\documentclass{edp-jp4}
\usepackage{graphicx}
\begin{document}

\title
{
Scattering in tight atom waveguides
}
\author{M. G. Moore}
\address
{
Department of Physics \& Astronomy, Ohio University, Athens, OH 45701, 
USA
}
\sameaddress{4}
\author{T. Bergeman}
\address
{
Department of Physics \& Astronomy,
SUNY, Stony Brook, NY
11794-3800, USA
}
\sameaddress{4}
\author{M. Olshanii}
\address
{
Department of Physics \& Astronomy,
University of Southern California,
Los Angeles, CA 90089-0484, USA
}
\secondaddress
{
Institute for Theoretical Atomic and Molecular Physics,
Harvard-Smithsonian Center for Astrophysics,
Cambridge, MA 02138, USA
}
%
%
\maketitle
\begin{abstract}
Using the Theory of Scattering in Restricted Geometries developed by A. 
Lupu-Sax  as a
starting point, we present a comprehensive multi-channel theory of 
atom-atom scattering in
tight atom waveguides.
\end{abstract}
\maketitle
\tableofcontents

\subsection*{Conventions and notations}
%
{\small
\subsubsection*{\it Square root conventions}
Throughout the text we will be using two complementary conventions 
for the complex square root function: the usual convention
$\sqrt{z}$ defined as 
\begin{eqnarray*}
\sqrt{|z|e^{i\phi}} = \sqrt{|z|} e^{i\phi/2} \quad\quad 0 \le \phi < 2\pi
\quad,
\end{eqnarray*}
and a non-traditional $\sqrt[\downarrow]{z}$ one defined as 
\begin{eqnarray*} 
\sqrt[\downarrow]{|z|e^{i\phi}} 
= \sqrt{|z|} e^{i\phi/2} \quad\quad -2\pi < \phi \le 0 
\quad.   
\end{eqnarray*}  

\subsubsection*{\it Some notations}
\begin{tabular}{lcr}
$\hat{G}_{\hat{H}}(E)$  && Green's function of a Hamiltonian $\hat{H}$
\\
$\hat{T}_{\hat{H},\hat{V}}(E)$ && T-matrix of a perturbation $\hat{V}$ to
a Hamiltonian $\hat{H}$
\\
$\hat{G}(E)$ && Green's function of the harmonic waveguide with no scatterer 
present
\\
$\chi_{\hat{H}}(E)$
&&
regular part of the Green's function of a Hamiltonian $\hat{H}$ at the origin
\\
$\chi(E)$
&&
regular part of the harmonic waveguide Green's function at the origin
\\
${\cal E}=E/2\hbar\omega_\perp-1/2$ && rescaled and renormalized energy
\\
$a_\perp=\sqrt{\hbar/\mu\omega_\perp}$
&&
transverse harmonic oscillator length
\\
$|nm\rangle$ && eigenstates of the two-dimensional harmonic oscillator
\\
$a$ && three-dimensional scattering length
\\
$g=2\pi\hbar^2 a/\mu$
&&
three-dimensional coupling constant
\\
$a_{1D}$ && one-dimensional scattering length
\\
$g_{1D} = -\hbar^2/\mu a_{1D}$ && one-dimensional coupling constant
\end{tabular}

}
\section{Introduction}
Atom waveguides are a fundamental component of Atom Optics, and are 
expected to play an
important role in Atom Interferometry and Quantum Computing applications. 
To ensure proper
coherence as atomic beams propagate through the waveguides, effort should 
be made to avoid
decoherence-inducing mechanisms such as collisional losses and collisional 
phase shifts.
This clearly requires a detailed understanding of the effects of 
quasi-one-dimensional
confinement on atom-atom collisions. In particular, we will see that in the 
few-mode regime,
necessary for coherent propagation, the free-space estimates of the 
collisional effects are
no longer valid and a waveguide-specific theory is needed.

Such one-dimensional interacting atomic quantum gases have recently been 
attracting
significant theoretical and experimental interest, having become accessible 
via adiabatic
transfer from atomic Bose condensates to highly elongated tight cigar-shape 
traps. Here the
question of {\it effective one-dimensional coupling constants} becomes 
important, as they
play a crucial role in determining whether or not the ground state posses 
long-range order
(coherence). In the regime of tight confinement it is the virtual 
excitation of the
transverse modes which play a crucial role, leading to the 
confinement-induced
renormalization of interatomic interactions \cite{Tonks_PRL,Tom} (see also 
the
two-dimensional analog of such a renormalization described in 
\cite{Gora_2D_tight} and the
fermionic p-wave analog in \cite{Doerty_fermions}).

In this paper we present a comprehensive theory of atom-atom scattering in 
atom waveguides.
Our theory correctly takes into account both quantization of the transverse 
motion and the
transverse renormalization of the collisional strength, otherwise 
inaccessible by the
free-space scattering theory.

From the formal point of view the problem reduces to a Schr\"odinger 
equation for two atoms
in a harmonic waveguide. In this paper we concentrate on the universal 
properties of the
waveguide scattering which are governed uniquely by the scattering length. 
The Theory of
Scattering in Restricted Geometries developed by A. Lupu-Sax 
\cite{Lupu-Sax} allows us to
describe these properties without invoking the full interaction potential, 
which is very
often unknown.

\section{Formulation of the scattering problem}
\label{subsec:formulation}

We begin from the Hamiltonian for two atoms under transverse harmonic 
confinement and
subject to an arbitrary interaction potential
\begin{eqnarray}
\label{H2atoms}
\hat{H}_2&=&-\frac{\hbar^2}{2m_1}\nabla^2_1-\frac{\hbar^2}{2m_2}\nabla^2_2
    +\frac{1}{2}m_1\omega_\perp^2{\bf r}^2_{1\perp}
    +\frac{1}{2}m_2\omega_\perp^2{\bf r}^2_{2\perp}\nonumber\\
    &+&V({\bf r}_1-{\bf r}_2)
\end{eqnarray}
where $m_1$ and $m_2$ are the atomic masses, $\omega_\perp$ is the 
transverse trap
frequency, and $\nabla^2_i$ and ${\bf r}_{i\perp}$ are the Laplacian and 
radial coordinate
of the $i^{th}$ atom, respectively. This Hamiltonian is separable in 
relative and
center-of-mass coordinates ${\bf R}=(m_1{\bf r}_1+m_2{\bf r}_2)/M$, ${\bf 
r}={\bf r}_1-{\bf
r}_2$, $M=m_1+m_2$ being the total mass, yielding 
$\hat{H}_2=\hat{H}_{rel}+\hat{H}_{COM}$,
where
\begin{equation}
\label{Hrel}
    \hat{H}_{rel}=-\frac{\hbar^2}{2\mu}\nabla^2_{\bf 
r}+\frac{1}{2}\mu\omega_\perp^2{\bf r}^2_\perp
    +V({\bf r}),
\end{equation}
and
\begin{equation}
\label{HCOM}
    \hat{H}_{COM}=-\frac{\hbar^2}{2M}\nabla^2_{\bf 
R}+\frac{1}{2}M\omega_\perp^2{\bf R}^2_\perp,
\end{equation}
where $\mu=m_1m_2/(m_1+m_2)$ is the reduced mass, and ${\bf r}_\perp$ and 
${\bf R}_\perp$
are the relative and center-of-mass radial coordinates, respectively. The 
center-of-mass
Hamiltonian is that of a simple harmonic oscillator whose solution is 
known, hence we focus
only on the relative motion of the two particles. This reduces the problem 
to a single
particle of mass $\mu$, subject to transverse harmonic confinement, which 
is scattered by
an external potential $V({\bf r})$. The central equation which must be 
solved is therefore
Schr\"odinger's equation for the state of relative motion of the two atoms
\begin{equation}
\label{HpsiE}
    \left[E-\hat{H}-\hat{V}\right]|\psi(E)\rangle =0
\end{equation}
where $\hat{H}=\hat{H}_{rel}$ and $\hat{V}$ is the interatomic potential. 
Determining the
eigenstates of this Hamiltonian, in particular within the s-wave scattering 
approximation
for $V({\bf r})$, is the central goal of this paper.

\section{Green's function and T-matrix formalism}

\subsection{Definitions and theorems}
\label{subsec:definitions}

\begin{figure}
\begin{center}
\includegraphics[width=11cm]{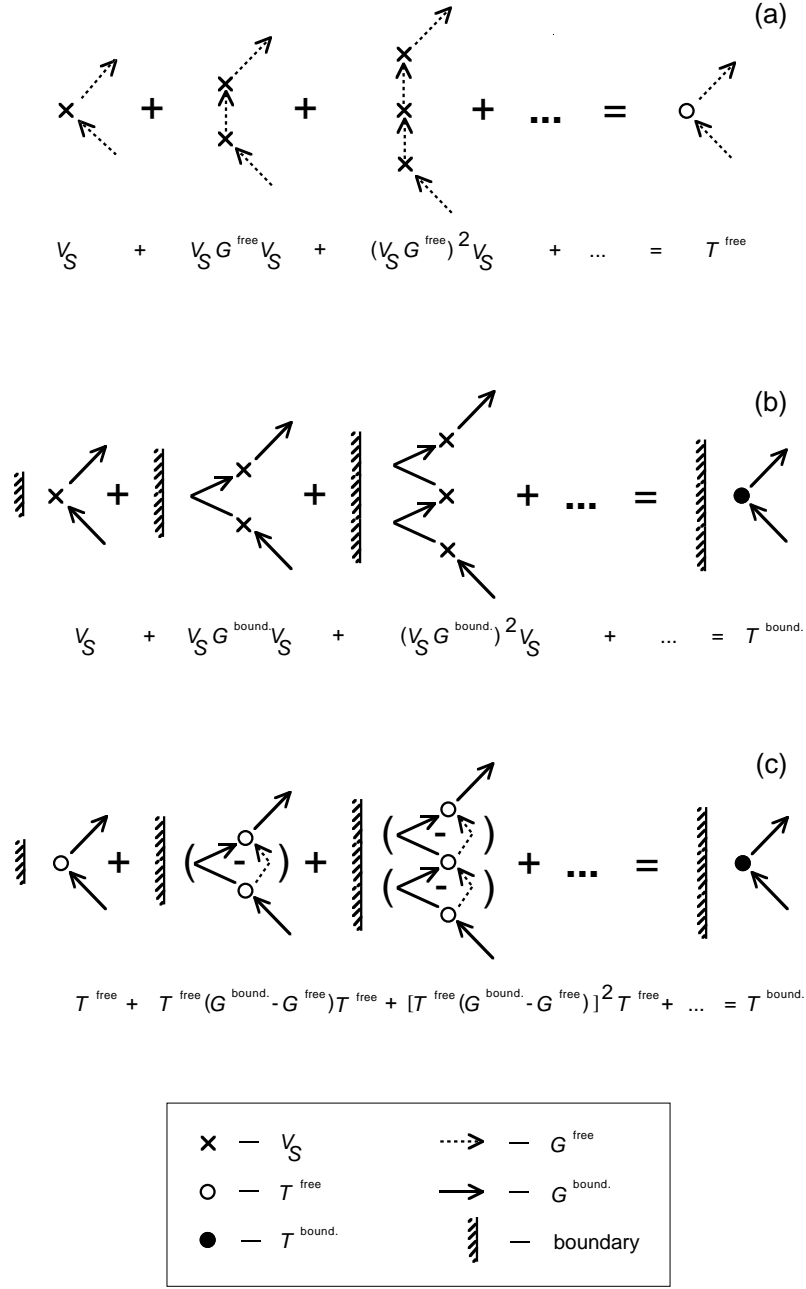}
\end{center}
\caption{
An artist view on the Lupu-Sax theorem. Here
$G^{free} = \hat{G}_{\hat{H}_0}(E)$, 
$G^{bound.} = \hat{G}_{\hat{H}_0+\hat{U}^{bound.}}(E)$,
$T^{free} = \hat{T}_{\hat{H}_0,\hat{V}_{s}}(E)$,
$T^{bound.} 
= \hat{T}_{\hat{H}_0 + \hat{U}^{bound.},\hat{V}_{s}}(E)$. In our 
problem $\hat{H}_0$ corresponds to the kinetic energy, $\hat{U}^{bound.}$
is the waveguide potential, and $\hat{V}_{s}$ is the interatomic interaction
potential.
}
\label{fig:scattering_theory}
\end{figure}

In this section briefly review the T-matrix formulation of scattering 
theory, which
provides a convenient framework for approaching the present problem. Let us 
first introduce
the retarded Green's function for a system with the Hamiltonian $\hat{H}$ 
and energy $E$
\begin{equation}
\label{GEH}
    \hat{G}_{\hat{H}}(E)
    = \lim_{\epsilon\to 0^+}(E + i\epsilon - \hat{H})^{-1}.
\end{equation}
We then define the T-matrix at energy $E$ of the scatter $\hat{V}$ in the 
presence of the
background Hamiltonian $\hat{H}$ (Fig.\ref{fig:scattering_theory}a,b) in 
the usual manner as
\begin{eqnarray}
\label{TEofVG}
\hat{T}_{\hat{H},\hat{V}}(E)&=&\left[1-\hat{V}\hat{G}_{\hat{H}}(E)\right]^{-1}
    \hat{V}\nonumber\\
    &=&\sum_{n=0}^\infty \left[\hat{V}\hat{G}_{\hat{H}}(E)\right]^n\hat{V},
\end{eqnarray}
the summation form being valid provided that there are no difficulties with 
convergence.

Two relations on which we will rely heavily are the Lippman-Schwinger 
relation
\begin{equation}
\label{LippSchwing}
    \hat{G}_{\hat{H}+\hat{V}}(E)=\hat{G}_{\hat{H}}(E)
    +\hat{G}_{\hat{H}}(E)\hat{T}_{\hat{H},\hat{V}}(E)\hat{G}_{\hat{H}}(E),
\end{equation}
which relates the full Green's function of the system $\hat{H}+\hat{V}$ to 
the unperturbed
Green's function $\hat{G}_{\hat{H}}(E)$ and the T-matrix, and the Lupu-Sax 
formula
(Fig.\ref{fig:scattering_theory}c)
\begin{equation}
\label{LupuSax}
    \hat{T}_{\hat{H},\hat{V}}(E)=
    \left[1-\hat{T}_{\hat{H}',\hat{V}}(E)
    \left[\hat{G}_{\hat{H}}(E)-\hat{G}_{\hat{H}'}(E)\right]
    \right]^{-1}\hat{T}_{\hat{H}',\hat{V}}(E),
\end{equation}
which relates the T-matrix of the scatter $\hat{V}$ in the background 
Hamiltonian $\hat{H}$
(Fig.\ref{fig:scattering_theory}b)
to the T-matrix for the same scatter but in a different background 
Hamiltonian $\hat{H}'$
(Fig.\ref{fig:scattering_theory}a).
Derivations for these expressions are given in Appendices \ref{LippSchwApp} 
and
\ref{LupSaxApp}, respectively.

\subsection{Scattering theory}
\label{subsec:scattering_theory}
In the continuous part of the spectrum of the total Hamiltonian 
$\hat{H}+\hat{V}$
its eigenstates can be expressed as a sum of
an incident and a scattered wave according to
\begin{equation}
\label{psispis0}
    |\psi(E)\rangle=|\psi_0(E)\rangle+|\psi_s(E)\rangle,
\end{equation}
where $|\psi_0(E)\rangle$, the `incident' state vector, satisfies
\begin{equation}
\label{Gpsi0}
   \hat{G}^{-1}_{\hat{H}}(E)|\psi_0(E)\rangle=0,
\end{equation}
we can then express the Schr\'odinger equation for the total system as
\begin{equation}
\label{GVpsi}
    \left[\hat{G}^{-1}_{\hat{H}}(E)-\hat{V}\right]
    \left(|\psi_0(E)\rangle+|\psi_s(E)\rangle\right)=0.
\end{equation}
This equation is readily solved for the scattered wave in terms of the 
unperturbed Green's
function and the T-matrix (see Appendix \ref{GVTApp}), yielding
\begin{equation}
\label{TEHV}
|\psi_s(E)\rangle=\hat{G}_{\hat{H}}(E)\hat{T}_{\hat{H},\hat{V}}(E)|\psi_0(E
)\rangle,
\end{equation}
which will serve as the basis for our treatment of the present scattering 
problem.

\subsection{Bound T-matrix as a generator for bound state energies}
\label{subsec:bound_T-matrix}

It follows from Eq. (\ref{GEH}) that $\hat{G}_{\hat{H}}(E)$ is diagonal in 
energy
representation and therefore takes the form
\begin{equation}
\label{GEbasis}
    \hat{G}_{\hat{H}}(E)=
    \sum_n\frac{|E_n\rangle\langle E_n|}{E-E_n}
    +\sum_j\int^{E_{u,j}}_{E_{l,j}} dE'\
    \frac{|E',j\rangle\langle E',j|}{E-E'+i\epsilon},
\end{equation}
where the first term on the r.h.s. sums over the discrete portion of the 
spectrum of
$\hat{H}$ and the second term sums over all continuous bands of the 
spectrum. The limit
$\epsilon\to 0^+$ is implied. From this expression we see that in the 
discrete
(bound-state) part of the spectrum the poles of the Green's function 
correspond to the
bound state energies.

Consider now the particular case where the background Green's function
$\hat{G}_{\hat{H}}(E)$ has only a continuous spectrum bounded from below by 
$E=0$, i.e.
\begin{equation}
\label{GHtoy}
    \hat{G}_{\hat{H}}(E)=\int^\infty_0 dE'\ \frac{|E'\rangle\langle 
E'|}{E-E'+i\epsilon}.
\end{equation}
Now consider a system consisting of this background Hamiltonian $\hat{H}$ 
plus a scattering
potential $\hat{V}$, described by the T-matrix 
$\hat{T}_{\hat{H},\hat{V}}(E)$. Inserting
equation (\ref{GHtoy}) into the Lippman-Schwinger Equation 
(\ref{LippSchwing}) for the total
Green's function gives
\begin{equation}
\label{GHVtoy}
    \hat{G}_{\hat{H}+\hat{V}}(E)
    =\int^\infty_0 dE'\ \frac{|E'\rangle\langle E'|}{E-E'+i\epsilon}
    +\int^\infty_0\int^\infty_0 dE' dE''
    \frac{|E'\rangle\langle 
E'|\hat{T}_{\hat{H},\hat{V}}(E)|E''\rangle\langle E''|}
    {(E-E'+i\epsilon)(E-E''+i\epsilon)}
\end{equation}
If a bound state of the combined system $\hat{H}+\hat{V}$ having energy 
$E_n<0$ exists,
then the full Green's function must take the form
\begin{equation}
\label{Gfullnerres}
    \hat{G}_{\hat{H}+\hat{V}}(E)\stackrel{E \approx E_n}{\approx}
    \frac{|E_n\rangle\langle E_n|}{E-E_n}.
\end{equation}
Since the background Green's function has no pole at $E_n$, it follows that 
the T-matrix
itself must have a singularity at $E=E_n$. Finding the poles of the 
T-matrix below the lower
bound of the spectrum of the background Hamiltonian therefore gives a 
method for determining
the bound state energies of the system $\hat{H}+\hat{V}$.

\section{S-wave scattering regime: the reference T-matrix approach}
\label{swavepseudopot}

At first glance it may seem that finding the T-matrix is no easier than a 
direct solving of
the Schr\"odinger equation (\ref{HpsiE}). We will demonstrate, however, 
that the T-matrix
formulation allows for a self-consistent description of the low-energy part 
of the spectrum
that uses the {\it free-space} low-energy scattering properties of the 
interaction potential
as the {\it only} input. In addition the low-energy (s-wave) limit is 
isolated to a single
well-defined approximation without requiring the ad-hoc introduction of 
regularization via a
pseudo-potential. In this section we first outline this self-consistent 
low-energy
treatment. We then solve for the T-matrix using the standard Huang-Fermi 
pseudo-potential,
showing that the pseudo-potential reproduces the exact result in this 
situation.

Let the unperturbed Hamiltonian $\hat{H}$ be a Hamiltonian for a single 
nonrelativistic
particle in presence of a trapping potential $U$:
\begin{eqnarray}
\label{Hgeneral}
    \langle{\bf r}|\hat{H}|\psi\rangle
    =\left[ -\frac{\hbar^2 \nabla^2_{{\bf r}}}{2\mu} + U({\bf 
r})\right]\langle{\bf r}|\psi\rangle \quad ,
\end{eqnarray}
Assume also that the particle is` perturbed' by a scatterer given by
\begin{eqnarray}
\label{Vgeneral}
    \langle{\bf r}|\hat{V}|\psi\rangle
    =V({\bf r})\langle{\bf r}|\psi\rangle \quad
\end{eqnarray}
localized around ${\bf r} = {\bf 0}$. In what follows we will derive a {\it 
low-energy
approximation} for the T-matrix of the scatterer $V$ in presence of 
$\hat{H}$. It is
important to note that, by definition, the T-matrix acts only on 
eigenstates of the
unperturbed Hamiltonian, which we can safely assume to be regular 
everywhere (this is of
course a constraint on the properties of the unperturbed Hamiltonian). In 
this case the
zero-range s-wave scattering limit does not require any regularization of 
the T-matrix. By
making use of the Lupu-Sax formula (\ref{LupuSax}), we first derive the 
correct form of the
T-matrix in the low-energy s-wave regime without the introduction of a 
regularized
pseudo-potential. In the following section, however, we will see that the 
results we obtain
are in agreement with the standard Huang-Fermi pseudopotential approach to 
s-wave
scattering.

We begin our derivation by first specifying a `reference' background 
Hamiltonian $\hat{H}'$
as
\begin{eqnarray}
\label{H'}
    \langle{\bf r}|\hat{H}'|\psi\rangle
    =\left[-\frac{\hbar^2 \Delta_{{\bf r}}}{2\mu}+E\right]\langle{\bf 
r}|\psi\rangle.
\end{eqnarray}
This Hamiltonian is that of a free particle, but with an explicit energy 
dependence included
so that the eigenstates have zero wavelength at all energies. We note that 
this reference
Hamiltonian agrees with the free-space Hamiltonian in the zero-energy 
limit. While this
Hamiltonian may seem strange, it is a valid reference Hamiltonian which 
turns out to be
useful because the resulting T-matrix is energy independent for any 
scattering potential.
The Green's function for this Hamiltonian is given by
\begin{eqnarray}
\label{G'}
     \langle {\bf r} | \hat{G}_{\hat{H}'}(E) | {\bf r}' \rangle =
     -\frac{\mu}{2\pi \hbar^2}
    \frac{1}{|{\bf r}-{\bf r}^{\prime}|},
\end{eqnarray}
as can be verified by direct substitution into $[E-\hat{H}']\,
\hat{G}_{\hat{H}'}(E)=\hat{I}$. In turn the $T$-matrix of the interaction 
potential $V$ in
presence of $\hat{H}'$ is independent of energy and can therefore be 
expressed as
\begin{equation} \label{T'}
     \langle {\bf r}|\hat{T}_{\hat{H}',\hat{V}}(E)|{\bf r}'\rangle
    =g D({\bf r},{\bf r}'),
\end{equation}
where the kernel $D$ is defined as normalized to unity,
\begin{equation}
\label{Dnorm}
    \int d{\bf r} d{\bf r}' \, D({\bf r},{\bf r}') = 1.
\end{equation}
The normalization coefficient $g$ is then related to the 3-dimensional 
scattering length $a$
according to
\begin{equation}
    g=\frac{2\pi\hbar^2 a}{\mu},
\label{g}
\end{equation}
a relationship which is derived explicitly in Appendix \ref{gaApp}.

Imagine that the kernel $D({\bf r},{\bf r}')$ is well localized within some 
radius $R$. In
perturbative expansions at low energies this kernel only participates in 
convolutions with
slow (as compare to $R$) functions, in which case it can be approximated by 
a
$\delta$-function,
\begin{equation}
\label{Ddelta}
    D({\bf r},{\bf r}') \approx \delta({\bf r}) \delta({\bf r}').
\end{equation}
This straightforward approximation is the key to the s-wave scattering 
approximation. As we
demonstrate in detail in Appendix \ref{lowkApp}, this effectively replaces 
the exact
reference T-matrix by its long-wavelength limit, so that the reference 
T-matrix assumes the
form
\begin{equation}
\label{T'delta}
    \langle {\bf r}|\hat{T}_{\hat{H}',\hat{V}}(E)|{\bf r}'\rangle
    \stackrel{k,k^{\prime} \ll 1/R}{\approx} g \delta({\bf r})\delta({\bf r}'),
\end{equation}
which is equivalent to
\begin{equation}
\label{TH'E}
    \hat{T}_{\hat{H}',\hat{V}}(E)=g|0\rangle\langle 0|,
\end{equation}
where $|0\rangle$ is the position eigenstate corresponding to the location 
of the scatterer.
In expression (\ref{T'delta}) $k$ and $k'$ refer to the wavevectors of any 
matrices which
multiply the T-matrix from the left and right, respectively.

If we now substitute the above expression for the reference T-matrix into 
the Lupu-Sax
formula (\ref{LupuSax}) for the T-matrix under the background Hamiltonian 
$\hat{H}$ we
arrive at
\begin{eqnarray}
\label{expansion_1}
    \hat{T}_{\hat{H},\hat{V}}(E)
    &=&\sum_{n=0}^\infty \left[g|0\rangle\langle 
0|\hat{G}_{\hat{H}'}(E)\right]^ng|0\rangle\langle 0|\nonumber\\
    &=&
    \left[1-g\langle 0|\hat{G}_{\hat{H}}(E)|0\rangle+g\langle 
0|\hat{G}_{\hat{H}'}(E)|0\rangle\right]^{-1}
    g|0\rangle\langle 0|
    .\nonumber\\
\end{eqnarray}
Making use of Eq. (\ref{G'}), we introduce the function $\chi({\cal E})$, 
defined as
\begin{eqnarray}
\label{chi}
    \chi(E)=\lim_{{\bf r}\to 0}\left[\langle{\bf 
r}|\hat{G}_{\hat{H}}(E)|0\rangle
    +\frac{\mu}{2\pi\hbar^2|{\bf r}|}\right],
\end{eqnarray}
from which we obtain the following simple expression for the T-matrix of 
the scatterer
$\hat{V}$ in presence of the trap:
\begin{equation}
\label{Tzerorange}
    \langle {\bf r} | \hat{T}_{\hat{H},\hat{V}}(E) | \psi \rangle
    \stackrel{E \ll \hbar^2/\mu R^2}{\approx}
    \frac{g\delta({\bf r})}{1-g\chi(E)}
    \langle {\bf r}
    | \psi \rangle.
\end{equation}
As we will explicitly demonstrate for the case of transverse harmonic 
confinement, the
singularity in bound Green's function is the same as that in the free-space 
Green's
function. Hence, $\chi(E)$ is the value of the regular part of the bound 
Green's function at
the origin.

\section{The function $\chi(E)$}
\label{functionchi}

In this section we derive the bound Green's function of the waveguide, from 
which we obtain
an analytic expression for regular part at the origin $\chi(E)$, which 
gives the energy
dependence of the bound T-matrix. We begin by reviewing the eigenstates of 
the waveguide
potential, which we then use as a basis for expanding the bound Green's 
function and
obtaining $\chi(E)$.

\subsection{Eigenstates of the waveguide Hamiltonian}

The Hamiltonian for the relative motion of two atoms in a harmonic 
waveguide contains two
parts, the longitudinal free Hamiltonian $\hat{H}_z$ and the transverse 
confinement
Hamiltonian $\hat{H}_\perp$,
\begin{eqnarray}
\label{H}
    \hat{H}=\hat{H}_z+\hat{H}_\perp,
\end{eqnarray}
where
\begin{equation}
    \langle{\bf r}|\hat{H}_z|\psi\rangle
    =-\frac{\hbar^2}{2\mu}\frac{\partial^2}{\partial z^2}\langle{\bf 
r}|\psi\rangle,
\label{Hz}
\end{equation}
and
\begin{equation}
    \langle{\bf r}|\hat{H}_\perp|\psi\rangle
    =\left[-\frac{\hbar^2}{2\mu}\left(\frac{\partial^2}{\partial \rho^2}
    +\frac{1}{\rho}\frac{\partial}{\partial\rho}
    +\frac{1}{\rho^2}\frac{\partial^2}{\partial\phi^2}\right)
    +\frac{1}{2}\mu\omega^2_\perp\rho^2\right]\langle{\bf r}|\psi\rangle.
\label{Hperp}
\end{equation}
The eigenstates of the transverse Hamiltonian, denoted by $|nm\rangle$, are 
well known and
satisfy both
\begin{equation}
    \hat{H}_\perp|n,m\rangle=\hbar\omega_\perp(2n+|m|+1)|nm\rangle,
\label{Enm}
\end{equation}
as well as
\begin{equation}
    \hat{L}_z|nm\rangle=\hbar m|nm\rangle,
\label{Lnm}
\end{equation}
where $\hat{L}_z$ is the operator for angular momentum along the z-axis. We 
note that in
this representation the radial and azimuthal quantum numbers $n$ and $m$ 
independently
assume the values
\begin{equation}
    n=0,1,2,\ldots,\infty,
\label{n}
\end{equation}
and
\begin{equation}
    m=0,\pm 1,\pm 2,\ldots,\pm\infty.
\label{nm}
\end{equation}
The eigenfunctions are given in r-representation by
\begin{equation}
    \langle\rho\phi|nm\rangle=\left[\frac{\pi 
a^2_\perp(n+|m|)!}{n!}\right]^{-1/2}
    e^{-{1\over 
2}(\rho/a_\perp)^2}(\rho/a_\perp)^{|m|}e^{i m\phi}L^{[|m|]}_n(\rho^2/a^2_\perp),
\label{rhophinm}
\end{equation}
where
\begin{equation}
    a_\perp=\sqrt{\frac{\hbar}{\mu\omega_\perp}}
\label{aperp}
\end{equation}
is the transverse harmonic oscillator length. Lastly we note that the value 
of
$|n,m\rangle$ at the origin is given by
\begin{equation}
    \langle 0|nm\rangle=\frac{\delta_{m,0}}{\sqrt{\pi}a_\perp},
\label{0phinm}
\end{equation}
which is independent of $n$.

\subsection{The Green's function for the relative motion
of two particles in a harmonic waveguide}

The bound Green's function is the solution to the equation
\begin{equation}
    \left[E-\hat{H}_z-\hat{H}_\perp\right]\hat{G}(E)=\hat{I},
\label{GEeq}
\end{equation}
where $\hat{I}$ is the identity matrix and for simplicity we have taken
$\hat{G}_{\hat{H}}(E)\leftrightarrow\hat{G}(E)$. Expanding this equation 
onto the set of
states
\begin{equation}
    |nmz\rangle=|nm\rangle\otimes|z\rangle,
\label{nmz}
\end{equation}
where $|nm\rangle$ is the eigenstate of $\hat{H}_\perp$ given by 
(\ref{rhophinm}) and
$|z\rangle$ is the eigenstate of $\hat{z}$, satisfying $\langle 
z|z'\rangle=\delta(z-z')$,
then gives
\begin{eqnarray}
\left[E-\hbar\omega_\perp(2n+|m|+1)+\frac{\hbar^2}{2\mu}\frac{\partial^2}
{\partial z^2}
    \right]\langle nmz|\hat{G}(E)|n'm'z'\rangle
    &=&\langle nmz|n'm'z'\rangle\nonumber\\
    &=&\delta_{n,n'}\delta_{m,m'}\delta(z-z').
\label{GEnmzeq}
\end{eqnarray}
We proceed by making the Ansatz for the Green's function
\begin{eqnarray}
    \langle nmz|\hat{G}(E)|n'm'z'\rangle&=&\delta_{n,n'}\delta_{m,m'}
    \alpha_{nm}e^{i\gamma_{nm}|z-z'|}\nonumber\\
    &=&\delta_{n,n'}\delta_{m,m'}
\alpha_{nm}\left[e^{i\gamma_{nm}(z-z')}\Theta(z-z')+e^{-i\gamma_{nm}(z-z')}
\Theta(z'-z)\right],
\label{ansatz}
\end{eqnarray}
where $\Theta(z)$ is the Heavyside step-function. Differentiating 
(\ref{ansatz}) twice
with respect to $z$ gives
\begin{equation}
    \frac{\partial^2}{\partial z^2}\langle nmz|\hat{G}(E)|n'm'z'\rangle
    =-\gamma^2_{nm}\langle 
nmz|\hat{G}(E)|n'm'z'\rangle+2i\gamma_{nm}\alpha_{nm}
    \langle nmz|n'm'z'\rangle,
\label{d2Gdz2}
\end{equation}
so the Eq. (\ref{GEnmzeq}) becomes
\begin{equation}
\left[E-\hbar\omega_\perp(2n+|m|+1)-\frac{\hbar^2\gamma^2_{nm}}{2\mu}\right]
    \langle nmz|\hat{G}(E)|n'm'z'\rangle
    +i\frac{\hbar^2\gamma_{nm}\alpha_{nm}}{\mu}\langle nmz|n'm'z'\rangle
    =\langle nmz|n'm'z'\rangle.
\label{GEnmzeq2}
\end{equation}
This equation is satisfied provided that
\begin{equation}
\gamma^2_{nm}=\frac{2\mu}{\hbar^2}\left[E-\hbar\omega_\perp(2n+|m|+1)\right]
\label{gammanm}
\end{equation}
and
\begin{equation}
    \alpha_{nm}=-i\frac{\mu}{\hbar^2\gamma_{nm}}.
\label{alphanm}
\end{equation}
Equation (\ref{gammanm}) is quadratic, and therefore has in general two 
solutions. We can
determine which solution is needed, however, from the causal nature of the 
retarded Green's
function. We should choose the solutions which propagate outwards from 
$z=z'$.

By introducing the dimensionless energy
\begin{equation}
    {\cal E}=\frac{E}{2\hbar\omega_\perp}-\frac{1}{2},
\label{epsilon}
\end{equation}
and making use of the relation (\ref{aperp}), we find that retarded Green's 
function can be
expressed as
\begin{equation}
\label{nmzGnmz}
    \langle nmz|\hat{G}({\cal E})|n'm'z'\rangle
    =-i\frac{\mu a_\perp}{2\hbar^2}
    \frac{e^{i\frac{2}{a_\perp}\sqrt{{\cal E}-n-\frac{|m|}{2}}\quad|z-z'|}}
    {\sqrt{{\cal E}-n-\frac{|m|}{2}}}
    \delta_{n,n'}\delta_{m,m'}.
\end{equation}
With this expression for the bound Green's function we can 
now proceed to
compute the bound T-matrix via equation (\ref{Tzerorange}).

\subsection{The function $\chi({\cal E})$ for a point scatterer in
presence of a wave-guide}

In order to evaluate $\chi({\cal E})$ we must first compute the matrix 
element $\langle
{\bf r}|\hat{G}({\cal E})|0\rangle$. Expanding this matrix element in terms 
of transverse
eigenstates gives
\begin{equation}
   \langle{\bf r}|\hat{G}({\cal E})|0\rangle
    =\sum_{nmn'm'}\langle \rho\phi|nm\rangle
    \langle nmz|\hat{G}({\cal E})|n'm'0\rangle
    \langle n'm'0 |0\rangle,
\label{rG0}
\end{equation}
which, with the help of Eqs. (\ref{0phinm}) and (\ref{nmzGnmz}), gives
\begin{equation}
    g\langle{\bf r}|\hat{G}({\cal E})|0\rangle
    =-\sqrt{\pi}a\sum^\infty_{n=0}\langle\rho\phi|n0\rangle
    \frac{e^{-\frac{2}{a_\perp}\sqrt[\downarrow]{n-{\cal E}}\ |z|}}
    {\sqrt[\downarrow]{n-{\cal E}}},
\label{rG02}
\end{equation}
where the modified square root $\sqrt[\downarrow]{}$ is defined in the 
``Conventions and notations'' table above. Now by 
inserting  Eqs.
(\ref{rG02}) and (\ref{g}) into Eq. (\ref{chi}) we arrive at the expression
\begin{equation}
    g\chi({\cal E})=\lim_{{\bf r}\to 0}
    \left[-\sqrt{\pi}a\sum_{n=0}^\infty\langle\rho\phi|n0\rangle\
    \frac{e^{-\frac{2}{a_\perp}\sqrt[\downarrow]{n - {\cal E}}|z|}}
    {\sqrt[\downarrow]{n-{\cal E}}}
+\frac{a}{|{\bf r}|}\right].
\label{gchiE2}
\end{equation}
While both terms in Eq. (\ref{gchiE2}) diverge in the limit ${\bf 
r}\to\infty$, we now
demonstrate that their difference remains finite and leads to an expression 
for $\chi({\cal
E})$ in terms of a generalized Zeta function.

First we assume, without a proof,  that the multi-variable 
limit in Eq. (\ref{gchiE2}) does exist, {\it i.e.} the 
directional single-variable limits ${\bf r} = s{\bf n}, s \to 0$
exist for all directions ${\bf n}$ and they are all equal to each other.
This assumption allows us to deal with the limit along the z-axis only:
\begin{equation}
    g\chi({\cal E})=-\frac{a}{a_\perp}\lim_{|z|\to 0}
    \left[
    \sum_{n=0}^\infty
    \frac{e^{-\frac{2}{a_\perp}\sqrt[\downarrow]{n - {\cal E}}|z|}}
    {\sqrt[\downarrow]{n-{\cal E}}}   
-\frac{a}{|z|}\right],
\label{gchiE10}
\end{equation}
where we have used the identity (\ref{0phinm}).
We proceed by first replacing the $a/|{\bf r}|$ term in Eq. (\ref{gchiE10}) 
with an integral
expression via the identity
\begin{equation}
    \frac{a}{|{\bf r}|}=\frac{a}{a_\perp}\int_{\cal E}^\infty dn\, 
\frac{e^{-\frac{2}{a_\perp}\sqrt{n-{\cal E}}|{\bf r}|}}
    {\sqrt{x-{\cal E}}}.
\label{intform}
\end{equation}
This allows us to write
\begin{equation} 
g\chi({\cal E})=-\frac{a}{a_\perp}
\lim_{|z|\to 0}
\lim_{N\to \infty}
\left[
    \sum_{n=0}^N
    \frac{e^{-\frac{2}{a_\perp}\sqrt[\downarrow]{n - {\cal E}}|z|}}
    {\sqrt[\downarrow]{n-{\cal E}}}
-
    \int_{\cal E}^N dn\,
    \frac{e^{-\frac{2}{a_\perp}\sqrt{n-{\cal E}}|{\bf r}|}}
    {\sqrt{n-{\cal E}}}
\right]    
,
\label{gchiE11}
\end{equation}
It is now tempting to interchange  the limit signs and thus get rid of the 
coordinate dependence. In order to be able to do that one have to prove the 
uniformity with respect to $|z|$ of the $N\to\infty$ convergence of the 
expression in the square brackets $\Xi(N,\,|z|)$, {\it i.e.} to prove that 
for every $\epsilon$ there exists $N^{\star}$, the same for all 
$|z|$, such that $\Xi(N,\,|z|) - \lim_{N\to \infty}\Xi(N,\,|z|) < \epsilon$
for all $N>N^{\star}$. Such a proof does exist: provided its length 
we do not exhibit it here. 

We arrive at the following expression for 
$\chi({\cal E})$:
\begin{equation} 
g\chi({\cal E})=-\frac{a}{a_\perp}
\lim_{N\to \infty}
\left[   
    \sum_{n=0}^N 
    \frac{1}
    {\sqrt[\downarrow]{n-{\cal E}}}
-
    2\,\sqrt{N-{\cal E}}
\right]   
.
\label{gchiE12}  
\end{equation}
One can now make use of the following theorem involving the Hurwitz Zeta 
function, an
analytic generalized Zeta function described in the mathematical literature 
\cite{Hurwitz},
\begin{eqnarray}
\label{zeta}
&&    \zeta(s,\alpha)=\lim_{N\to\infty}
\left[
\left(
    \sum_{n=0}^N \frac{1}{(n+\alpha)^{s}}
\right)
    - \frac{1}{1-s} \, \frac{1}{(N+\alpha)^{s-1}}
\right]
\\
\nonumber
&&\mbox{Re}(s)>0, \quad -2\pi < \arg(n+\alpha) \le 0.
\end{eqnarray}
In particular,
\begin{eqnarray} 
\label{zeta_1/2}   
&&    \zeta(1/2,\alpha)=\lim_{N\to\infty}
\left[   
\left(
    \sum_{n=0}^N \frac{1}{\sqrt[\downarrow]{n+\alpha}}
\right)  
    - 2\,\sqrt{N+\alpha} 
\right].  
\end{eqnarray}
While this expression, valid for any $N$, may be taken as a definition of 
the Hurwitz Zeta
function, it does not constitute an efficient method for computation. Most 
symbolic math
software packages will have efficient algorithms, however in using them 
care must be taken
with regards to the branch cut for fractional $s$. With this definition and 
taking $s=1/2$
we arrive at
\begin{equation}
    g\chi({\cal E})=-\frac{a}{a_\perp}\zeta(1/2,-{\cal E}).
\label{gchiEfinal}
\end{equation}

By substituting this expression into Eq. (\ref{Tzerorange}) we arrive at 
the final
expression for the long-wavelength T-matrix in the waveguide:
\begin{equation}
    \hat{T}({\cal E})=\frac{g|0\rangle\langle 
0|}{1+\frac{a}{a_\perp}\zeta(1/2,-{\cal E})}.
\label{Tfinal}
\end{equation}
We note that in the definition (\ref{zeta}) of the Hurwitz Zeta function, 
there is an
ambiguity in the sign of the square root. 

\section{Results}

\subsection{Multi-channel scattering amplitudes and transition rates}

From Equations (\ref{TEHV}) and (\ref{Tfinal}) we find that the scattered 
wavefunction
takes the form
\begin{equation}
\label{nmzpsis}
    \langle nmz|\psi_s({\cal E})\rangle
    =g\frac{\langle nmz|\hat{G}({\cal E})|0\rangle}
    {\left[1+\frac{a}{a_\perp}{\zeta}(1/2,-{\cal E})\right]}
    \langle 0|\psi_0({\cal E})\rangle.
\end{equation}
Now the matrix element $\langle nmz|\hat{G}({\cal E})|0\rangle$ can be 
determined by by
making use of Eqs. (\ref{nmzGnmz}) and (\ref{0phinm}), yielding
\begin{eqnarray}
\label{nmzGE0}
    \langle nmz|\hat{G}({\cal E})|0\rangle&=&\sum_{n'm'}\int dz'\
    \langle nmz|\hat{G}({\cal E})|n'm'z'\rangle
    \langle n'm'z'|0\rangle\nonumber\\
&=&-i\frac{\mu}{2\sqrt{\pi}\hbar^2}\delta_{m,0}\frac{e^{i\frac{2}{a_\perp}
\sqrt{{\cal E}-n}|z|}}
    {\sqrt{{\cal E}-n}}.
\end{eqnarray}
Inserting this expression into Eq. (\ref{nmzpsis}) then gives
\begin{equation}
\label{nmzpsis2}
    \langle nmz|\psi_s({\cal E})\rangle=
    -i\frac{\sqrt{\pi}a}{\left[1+\frac{a}{a_\perp}{\zeta}(1/2,-{\cal 
E})\right]}
    \delta_{m,0}\frac{e^{i\frac{2}{a_\perp}\sqrt{{\cal 
E}-n}|z|}}{\sqrt{{\cal E}-n}}
     \langle 0|\psi_0({\cal E})\rangle.
\end{equation}
Let us now assume that the incident wave has the longitudinal wave vector 
$k$ and the
transverse quantum numbers $n$ and $m$, according to
\begin{equation}
\label{nmzpsi0}
    \langle {\bf r}|\psi_0({\cal E})\rangle
    =\langle \rho\phi|nm\rangle e^{ikz},
\end{equation}
where the relation
\begin{equation}
\label{Ek}
    {\cal E}=\left(\frac{a_\perp k}{2}\right)^2+n+\frac{|m|}{2}
\end{equation}
gives the dependence of the scaled energy ${\cal E}$ on the incident 
wavevector $k$. This
incident wave is nonzero at the origin only for $m=0$, hence only incident 
waves with zero
angular momentum will scatter. The value of the $m=0$ incident wave at the 
origin
conveniently takes the $n$-independent value,
\begin{equation}
\label{0psi0E}
    \langle 0|\psi_0({\cal 
E})\rangle=\frac{\delta_{m,0}}{\sqrt{\pi}a_\perp}.
\end{equation}
Assuming henceforth $m=0$, we can now express Eq. (\ref{nmzpsis2}) as
\begin{equation}
\label{nmzpsis3}
    \langle n'm'z|\psi_s({\cal E})\rangle
    =-i\frac{\delta_{m',0}}
    {\left[\frac{a_\perp}{a}+{\zeta}(1/2,-{\cal E})\right]}
    \frac{e^{i\frac{2}{a_\perp}\sqrt{{\cal E}-n'}|z|}}{\sqrt{{\cal E}-n'}}.
\end{equation}
From this expression it follows that the full wavefunction of the relative 
motion takes the
form
\begin{equation}
\label{rpsiE}
    \langle {\bf r}|\psi({\cal E})\rangle
    =\sum_{n'=0}^\infty\langle\rho\phi|n'0\rangle
    \left[\delta_{n',n}e^{ikz}
    +f(k_{n'}\leftarrow k_n)_{n'\leftarrow n}
    e^{ik_{n'}|z|}\right].
\end{equation}
Here we have introduced the even-wave {\it transversely inelastic 
scattering amplitudes}
$f(k_{n'}\leftarrow k_n)_{n'\leftarrow n}$, given by
\begin{equation}
\label{fknn}
    f(k_{n'}\leftarrow k_n)_{n'\leftarrow n}=-\frac{2i}{a_\perp 
k_{n'}}\frac{1}
    {\left[\frac{a_\perp}{a}
    +{\zeta}(1/2,-\left(\frac{a_\perp k_n}{2}\right)^2-n)\right]},
\end{equation}
and the outgoing wave vector for the mode $|n'0\rangle$
\begin{equation}
\label{kn}
    k_{n'}=\frac{2}{a_\perp}\sqrt{\left(\frac{a_\perp 
k_n}{2}\right)^2+n-n'}\ ,
\end{equation}
from which the desired scattering probabilities can be computed.

One can now easily compute the elastic and inelastic transition 
probabilities for
collisions under transverse harmonic confinement. We begin by considering 
the asymptotic
forms of the total wavefunction given by Eq. (\ref{rpsiE}), which are given 
by
\begin{eqnarray}
    \lim_{z\to\infty}\langle n'0z|\psi({\cal E})\rangle
    &=&\delta_{n',n}e^{ikz}+\Theta\left[{\cal E}-n'\right]\
    f(k_{n'}\leftarrow k_n)_{n'\leftarrow n}
    e^{ik_{n'}z}\\
    \lim_{z\to-\infty}\langle n0z|\psi({\cal E})\rangle
    &=&\Theta\left[{\cal E}-n'\right]\
    f(k_{n'}\leftarrow k_n)_{n'\leftarrow n}e^{-ik_{n'}z}.
\end{eqnarray}
Because of energy conservation, an inelastic collision results in a change 
in the
longitudinal momentum $k\to k_n$, so that the introduction of inelastic 
transmission and
reflection coefficients must be based on conservation of total incident and 
outgoing
probability current. For longitudinal plane waves the probability current 
is given by the
amplitude squared times the velocity, which leads to the inelastic 
transmission and
reflection coefficients
\begin{eqnarray}
\label{TnmRnm}
    T(k_{n'}\leftarrow k_n)_{n'\leftarrow n}
    &=&\Theta\left[{\cal E}-n'\right]
    \sqrt{\frac{{\cal E}-n'}{{\cal E}-n}}
    \left|\delta_{n',n}+f(k_{n'}\leftarrow k_n)_{n'\leftarrow n}\right|^2\\
    R(k_{n'}\leftarrow k_n)_{n'\leftarrow n}&=&
    \Theta\left[{\cal E}-n'\right]\sqrt{\frac{{\cal E}-n'}{{\cal E}-n}}
    \left|f(k_{n'}\leftarrow k_n)_{n'\leftarrow n}\right|^2,
\end{eqnarray}
which are readily evaluated with the help of Eq. (\ref{fknn}). The sum of 
the transmission and reflection coefficients
of the particular channel gives the corresponding transition probability 
governing the population exchange between
the transverse vibrational levels.

Substituting the expression (\ref{fknn}) to the transmission and reflection 
coefficients above
we obtain the following set of kinetic coefficients:
\begin{eqnarray}
\label{kinetic}
&&W_{\hookrightarrow n}(k) = 1 - W_{\leftarrow n}(k)
\\
&&W_{\leftarrow n}(k) = \frac{2}{\sqrt{{\cal E}_{t}}}\,\frac{\eta({\cal 
E})-1/\sqrt{{\cal E}_{t}}}
                                      {[\frac{a}{a}_{\perp}+\zeta(1/2, 
1-\delta{\cal E})]^2 + \eta^2({\cal E}) }
\\
&&W_{n'\leftarrow n}(k) = \Theta\left[{\cal E}-n'\right]\,
\frac{2}{\sqrt{{\cal E}_{t}}}\, \frac{1}{\sqrt{{\cal E}-n'}}\,
\frac{1}
{[\frac{a}{a}_{\perp}+\zeta(1/2, 1-\delta{\cal E})]^2 + \eta^2({\cal E}) }
\\
&&T_{\hookrightarrow n}(k) = W_{\hookrightarrow n}(k) - R_{\hookrightarrow 
n}(k)
\\
&&R_{\hookrightarrow n}(k) = \frac{1}{{\cal E}_{t}}\,
\frac{1}
{[\frac{a}{a}_{\perp}+\zeta(1/2, 1-\delta{\cal E})]^2 + \eta^2({\cal E}) }
\\
&&T_{n'\leftarrow n}(k) = R_{n'\leftarrow n}(k) = W_{n'\leftarrow n}(k)/2
\end{eqnarray}
where $W_{\hookrightarrow n}(k)$ is the probability that after a collision 
of two particles with the
relative momentum $k$ and relative transverse excitation $n$ the particles 
will remain in the
same transverse state, this probability is a sum of transmission 
$T_{\hookrightarrow n}(k)$
and reflection $R_{\hookrightarrow n}(k)$ probabilities,
$W_{\leftarrow n}(k)$ is the total probability of changing the transverse
state, $W_{n'\leftarrow n}(k)$ is the probability of transition to a 
particular transverse channel,
the transmission $T_{n'\leftarrow n}(k)$
and reflection $R_{n'\leftarrow n}(k)$ probabilities in channel-changing 
collisions are equal to
each other. Here
\begin{eqnarray}
&&{\cal E} = n + {\cal E}_{t}
\\
&&{\cal E}_{t} = (ka_{\perp}/2)^2
\\
&&\delta{\cal E} = {\cal E} - [{\cal E}] \quad\quad 0\le \delta{\cal E} < 1
\\
&&\eta({\cal E}) = \sum_{n'=0}^{[{\cal E}]} \frac{1}{\sqrt{{\cal E}-n'}}
\end{eqnarray}
where $[\ldots]$ is the integral part sign.

\subsection{Single-channel scattering and effective one-dimensional 
interaction potential}
\label{subsect:1D_pot}

Let us now consider the special case of a single-channel scattering:
\begin{equation}
\label{singlech}
    0 \le {\cal E} < 1,\quad n_0=0
\end{equation}
In this case we have
\begin{equation}
\label{singchxi}
    g\chi({\cal E})=-\frac{a}{a_\perp}\zeta(1/2,-{\cal E})=
    -\frac{a}{a_\perp}\left[\zeta(1/2,1-{\cal E})+\frac{i}{\sqrt{\cal 
E}}\right].
\end{equation}
Using an alternative representation for $\zeta(1/2,1-{\cal E})$ 
\begin{equation}
\label{zeta1minusE}
    \zeta(1/2,1-{\cal E})=\lim_{N\to\infty}
    \left[\sum_{n=1}^N \frac{1}{\sqrt[\downarrow]{n-{\cal E}}}-2\sqrt{N}\right].
\end{equation}
and making use of the expansion in powers of ${\cal E}$,
\begin{eqnarray}
\label{powersE}
&&    \frac{1}{\sqrt{n-{\cal E}}}=\frac{1}{\sqrt{n}}
    +\sum_{j=1}^\infty\frac{(2j-1)!!}{2^jj!n^{j+1/2}}{\cal E}^j
\\
\nonumber
&&|{\cal E}| < 1, \quad n > 0,
\end{eqnarray}
allows us to write
\begin{equation}
\label{introLE}
    \zeta(1/2,1-{\cal E})=\zeta(1/2)+{\cal L}({\cal E}),
\end{equation}
where
\begin{equation}
\label{defLE}
    {\cal L}({\cal 
E})=\sum_{j=1}^\infty\frac{(2j-1)!!\zeta(j+1/2)}{2^jj!}{\cal E}^j,
\end{equation}
which clearly separates the zero energy limit from the finite energy 
corrections.

According to (\ref{fknn}) this leads to
\begin{equation}
\label{singchfk}
    f_e(k)=-\frac{1}
    {\left[1+ia_{1D}k-i\frac{a_\perp k}{2}{\cal 
L}\left(\frac{a_\perp^2k^2}{4}\right)\right]},
\end{equation}
where $f_e(k)=f(k_0\leftarrow k_0)_{0\leftarrow 0}$ is the even 
single-channel scattering
amplitude and
\begin{equation}
\label{a1D}
    a_{1D}=-\frac{a_\perp}{2}\left[\frac{a_\perp}{a}+\zeta(1/2)\right]
\end{equation}
with $\zeta(1/2)=-1.4603\ldots$, is the effective one-dimensional 
scattering length which
agrees with the result in \cite{Tonks_PRL}.

It is now tempting to introduce an effective one-dimensional interaction 
potential
in such a way that its scattering amplitude, introduced through the 
one-dimensional scattering solution
as
\begin{equation}
\psi(z) \stackrel{z\to \pm\infty}{=} \exp(ikz) + f_e(k)\exp(ik|z|) + f_o(k) 
\mbox{sign}(z) \exp(ik|z|)
\quad,
\end{equation}
matches (\ref{singchfk}), i.e.\ solve the
corresponding one-dimensional inverse scattering problem. It turns out that 
this problem is ill-posed
due to the presence of open transverse channels unaccessible within the 
one-dimensional model. Nevertheless
one may pose the following problem: find a one-dimensional potential, whose 
scattering amplitude reproduces the
exact one (\ref{singchfk}) with the relative error ${\cal O}(k^3)$. Such an 
object does exist, and it is represented by
a zero-range scatterer
\begin{equation}
v(z) = g_{1D}\delta(z)
\label{1D_pot}
\end{equation}
of a coupling strength $g_{1D} = -\hbar^2/\mu a_{1D}$.

Notice now the resonant behavior of $g_{1D}$ showing a Confinement Induced 
Resonance (CIR) at
$a = a_\perp/|\zeta(1/2)|$. The effect was recently interpreted in terms of
Feshbach resonance between ground and excited vibrational manifolds. The 
resonance
has been confirmed by numerical calculations with finite-range potentials,
at both two-body \cite{Tom} and many-body \cite{Doerty_CIR} level.

\subsection{Bound states}
\label{boundstates}

As it has been discussed in the section \ref{subsec:bound_T-matrix} the 
poles of the
full T-matrix of the problem (the T-matrix \ref{Tfinal} in our case) 
correspond to the bound states.
We get the following eigenvalue equation
\begin{equation}
   \zeta(1/2,-{\cal E}) = -\frac{a_\perp}{a}
\end{equation}
The detailed analysis, interpretation, and testing of the solutions against 
finite-range models
is presented in \cite{Tom}. It turns out that for any set of parameters 
there exists one and only
one bound state. For  small positive three-dimensional scattering length it 
converges to
the free-space three-dimensional bound state. For small negative scattering 
length
the bound state corresponds to the bound state of the one-dimensional 
potential (\ref{1D_pot}).

\section{Concluding remarks, related works, and open problems}
In this paper we have demonstrated that the low-energy free-space 
properties of a scatterer
are sufficient to describe its low-energy behavior in a non-free 
environment. The T-matrix
formalism and the Lupu-Sax connection formula in particular serve as a 
powerful bridge between the two.

Several related works ought to be mentioned.
Scattering of bosons in two-dimensional (planar) harmonic waveguides has been 
successfully treated in
\cite{Gora_2D_tight}. Fermions in a linear guide were considered in 
\cite{Doerty_fermions} via
the K-matrix formalism  and the
corresponding one-dimensional amplitudes were explicitly computed. In an 
$N$-body
setting $N-2$ particles can be interpreted as the background potential for 
a given pair
\cite{lambda}.

One would expect that the inverse scattering problem posed in section 
\ref{subsect:1D_pot}
can be solved with an accuracy higher than existing ${\cal O}(k^3)$. The 
solution
will allow to improve the accuracy of the many-body Monte-Carlo numerical 
models without
introducing the transverse dimensions, otherwise making the computation 
harder.

Shallow atomic guides with only a few transverse bound levels constitute a 
significant
challenge. Unlike for harmonic guides the separation of relative and 
transverse
degrees of freedom will be lifted leading to new collision channels. 
Presence
of continuum spectrum for virtual transverse excitations may significantly
enhance the renormalization effects or even lead to new resonances.


\Appendix
%
\section{Acknowledgments}
Authors are grateful to Adam Lupu-Sax, Rick Heller, Vanja Dunjko, Yvan 
Castin, and Dimitry Petrov
for enlightening discussions on the subject. This
work was supported by the National Science Foundation grant {\it 
PHY-0070333} (M.O.),
a grant from Office of Naval Research {\it N00014-03-1-0427} (T.B., M.0.),
and through the National Science Foundation grant for the Institute
for Theoretical Atomic and Molecular Physics at Harvard University and 
Smithsonian
Astrophysical Observatory.

\section{Appendices}

\subsection{The T-matrix, the potential and the background Green's function}
\label{GVTApp}

In this section we derive various useful relationships between the Green's 
function of a
given background Hamiltonian, $\hat{G}_{\hat{H}}(E)$, the scattering 
potential, $\hat{V}$,
and the T-matrix, $\hat{T}_{\hat{H},\hat{V}}(E)$, which connects the 
incident wavevector
$|\psi_0(E)\rangle $ with the scattered wave $|\psi_s(E)\rangle$ according 
to Eq.
(\ref{TEHV}). By making use of the definition of the T-matrix (\ref{TEHV}) 
and the fact
that the incident wave satisfies the unperturbed Hamiltonian (\ref{Gpsi0}), 
we can cast
Schr\"odinger's equation (\ref{GVpsi}) in the following form
\begin{equation}
\label{SchroT}
    \left[\hat{T}_{\hat{H},\hat{V}}(E)
    -\hat{V}\hat{G}_{\hat{H}}(E)\hat{T}_{\hat{H},\hat{V}}(E)-\hat{V}\right]
    |\psi_0(E)\rangle.
\end{equation}
Since this equation must be satisfied for any $|\psi_0(E)\rangle$ it 
follows that
\begin{equation}
\label{VGTeq}
    \hat{T}_{\hat{H},\hat{V}}(E)
    -\hat{V}\hat{G}_{\hat{H}}(E)\hat{T}_{\hat{H},\hat{V}}(E)-\hat{V}=0.
\end{equation}
This equation can be readily solved for either the T-matrix 
$\hat{T}_{\hat{H},\hat{V}}(E)$
or the potential $\hat{V}$. Solving first for 
$\hat{T}_{\hat{H},\hat{V}}(E)$ then gives
\begin{equation}
\label{TEofGV2}
\hat{T}_{\hat{H},\hat{V}}(E)=\left[1-\hat{V}\hat{G}_{\hat{H}}(E)\right]^{-1}
    \hat{V},
\end{equation}
which gives an expression for the T-matrix at energy $E$ in terms of the 
perturbation
$\hat{V}$ and the background Green's function $\hat{G}_{\hat{H}}(E)$. We 
can also solve Eq.
(\ref{VGTeq}) for $\hat{V}$, which yields
\begin{equation}
\label{VofGT}
    \hat{V}=\hat{T}_{\hat{H},\hat{V}}(E)
    \left[1+\hat{G}_{\hat{H}}(E)\hat{T}_{\hat{H},\hat{V}}(E)\right]^{-1},
\end{equation}
or by starting from the Hermite conjugate of Eq. (\ref{GVpsi}) we can 
similarly arrive at
the equivalent expression
\begin{equation}
\label{VofTG}
    \hat{V}=
    \left[1+\hat{T}_{\hat{H},\hat{V}}(E)\hat{G}_{\hat{H}}(E)\right]^{-1}
    \hat{T}_{\hat{H},\hat{V}}(E).
\end{equation}

\subsection{The Lippman-Schwinger equation}
\label{LippSchwApp}

In this section we derive the relation between the full Green's function of 
the Hamiltonian
$\hat{H}+\hat{V}$ in terms of the Green's function for the Hamiltonian 
$\hat{H}$ and the
T-matrix. The full Green's function can be defined by
\begin{equation}
\label{GEHV}
    \hat{G}^{-1}_{\hat{H}+\hat{V}}(E)=\hat{G}^{-1}_{\hat{H}}-\hat{V},
\end{equation}
which, after substituting Eq. (\ref{VofGT}) yields
\begin{equation}
\label{GEHT}
    \hat{G}^{-1}_{\hat{H}+\hat{V}}(E)=\hat{G}^{-1}_{\hat{H}}
    -\hat{T}_{\hat{H},\hat{V}}(E)\left[1+\hat{G}_{\hat{H}}(E)
    \hat{T}_{\hat{H},\hat{V}}(E)\right]^{-1}.
\end{equation}
Multiplying from the left by $\hat{G}_{\hat{H}+\hat{V}}(E)$ and then from 
the right by
$\left[1+\hat{G}_{\hat{H}}(E)\hat{T}_{\hat{H},\hat{V}}(E)\right]\hat{G}_{\hat{H}}(E)$ 
then
yields the Lippman-Schwinger relation
\begin{equation}
\label{LippSchwing1}
    \hat{G}_{\hat{H}+\hat{V}}(E)=\hat{G}_{\hat{H}}(E)
    +\hat{G}_{\hat{H}}(E)\hat{T}_{\hat{H},\hat{V}}(E)\hat{G}_{\hat{H}}(E)
\end{equation}
which relates the full Green's function to the `background' Green's 
function and T-matrix.

\subsection{The Lupu-Sax formula}
\label{LupSaxApp}

In many situations one may not have knowledge of the potential $\hat{V}$, 
but rather have
direct knowledge only of the T-matrix, $\hat{T}_{\hat{H}',\hat{V}}(E)$ with 
respect to some
background Hamiltonian $\hat{H}'$. If additional external fields are 
applied, then ideally
one would like an expression for the T-matrix of the same perturbation in 
the presence the
new background Hamiltonian $\hat{T}_{\hat{H},\hat{V}}(E)$.

We can derive the required expression directly from the relations 
(\ref{VofGT}) and
(\ref{VofTG}). Since $\hat{V}$ is equal to itself we can equate expression 
(\ref{VofGT})
for $\hat{V}$ in terms of $\hat{H}'$ to the equivalent expression 
(\ref{VofTG}) in terms of
$\hat{H}$, which yields the relation
\begin{equation}
\label{TGHTGH'}
    \left[1+\hat{T}_{\hat{H}',\hat{V}}(E)\hat{G}_{\hat{H}'}(E)\right]^{-1}
    \hat{T}_{\hat{H}',\hat{V}}(E)=\hat{T}_{\hat{H},\hat{V}}(E)
    \left[1+\hat{G}_{\hat{H}}(E)\hat{T}_{\hat{H},\hat{V}}(E)\right]^{-1}.
\end{equation}
Multiplying from the left by 
$1+\hat{T}_{\hat{H}',\hat{V}}(E)\hat{G}_{\hat{H}'}(E)$ and
from the right by $1+\hat{G}_{\hat{H}}(E)\hat{T}_{\hat{H},\hat{V}}(E)$ then 
gives
\begin{equation}
\label{TGHTGH'2}
    \hat{T}_{\hat{H},\hat{V}}(E)
    \left[1+\hat{G}_{\hat{H}}(E)\hat{T}_{\hat{H},\hat{V}}(E)\right
    ]=\left[1+\hat{T}_{\hat{H}',\hat{V}}(E)\hat{G}_{\hat{H}'}(E)\right]
    \hat{T}_{\hat{H},\hat{V}}(E).
\end{equation}
We note that in this equation the unknown operator 
$\hat{T}_{\hat{H},\hat{V}}(E)$ appears
linearly and always on the far-right side. Hence, solving for
$\hat{T}_{\hat{H},\hat{V}}(E)$ is straightforward and results in the 
Lupu-Sax formula
\cite{Lupu-Sax} relating the $T$-matrices of the same perturbation but in 
two different
background Hamiltonians $\hat{H}$ and $\hat{H}^{\prime}$:
\begin{equation}
\label{LupuSax2}
    \hat{T}_{\hat{H},\hat{V}}(E)=
    \left[1-\hat{T}_{\hat{H}',\hat{V}}(E)
    \left(\hat{G}_{\hat{H}}(E)-\hat{G}_{\hat{H}'}(E)\right)
    \right]^{-1}\hat{T}_{\hat{H}',\hat{V}}(E).
\end{equation}

\subsection{The free-space scattering length}
\label{gaApp}

In this section we will related the 3-dimensional scattering length to the 
normalization of
the kernel of the reference T-matrix (\ref{T'}). In Section 
\ref{swavepseudopot} we
introduced the reference Hamiltonian $\hat{H}'=\hat{H}_{free}+E$, where 
$\hat{H}_{free}$ is
the free-space Hamiltonian containing only a 3-dimensional kinetic energy 
term. We note that
for $E=0$ the reference Hamiltonian and the free-space Hamiltonian agree, 
hence we can use
the Green's function and T-matrix of $\hat{H}'$ to solve the free-space 
scattering problem
at zero energy. The free-space solution for scattering from the potential 
$\hat{V}$ is
therefore given at $E=0$ by
\begin{equation}
\label{psis0}
|\psi_s(0)\rangle=\hat{G}_{\hat{H}'}(0)\hat{T}_{\hat{H}',\hat{V}}(0)|\psi_0
(0)\rangle
\end{equation}
where $\hat{G}_{\hat{H}'}(0)$ is given by Eq. (\ref{G'}) and
$\hat{T}_{\hat{H}',\hat{V}}(0)$ by Eq. (\ref{T'}). Expanding this 
expression onto position
eigenstates and making use of (\ref{G'}) and (\ref{T'}) gives
\begin{equation}
\label{psis0r}
    \langle{\bf r}|\psi_s(0)\rangle
    =-\frac{\mu g}{2\pi\hbar^2} \int d{\bf r}' d{\bf r}'\
    \frac{D({\bf r}',{\bf r}'')}{|{\bf r}-{\bf r}'|}\langle{\bf 
r}''|\psi_0(0)\rangle.
\end{equation}
For zero energy the incident wave is given by $\langle{\bf 
r}|\psi_0(0)\rangle=1$, hence by
making the assumption that $D({\bf r}',{\bf r}'')$ is localized at ${\bf 
r}',{\bf
r}''\approx 0$ we find that the limit as ${\bf r}\to\infty$ is given by
\begin{eqnarray}
\label{rtoinftypsis0}
    \lim_{{\bf r}\to\infty}\langle{\bf r}|\psi_s(0)\rangle&=&
    -\frac{\mu g}{2\pi\hbar^2}\frac{1}{|{\bf r}|}
    \int d{\bf r}' d{\bf r}'' D({\bf r}',{\bf r}'')\nonumber\\
    &=& -\frac{\mu g}{2\pi\hbar^2}\frac{1}{|{\bf r}|}.
\end{eqnarray}
Because the potential is assumed to vanish as ${\bf r}\to\infty$, we know 
that the
asymptotic form of the scattered wavefunction is given by a spherical plane 
wave with zero
kinetic energy, characterized by the zero-energy scattering amplitude 
$f_s(0)$,
\begin{equation}
\label{asymptform}
    \lim_{{\bf r}\to\infty}\langle{\bf 
r}|\psi_s(0)\rangle=f_s(0)\frac{1}{|{\bf r}|}.
\end{equation}
Comparison with Eq. (\ref{rtoinftypsis0}) then shows that the zero-energy 
scattering
amplitude is
\begin{equation}
\label{fs0}
    f_s(E=0)=-\frac{\mu g}{2\pi\hbar^2}.
\end{equation}
The asymptotic form of the total wavefunction is then
\begin{equation}
\label{rtoinftypsi0}
    \lim_{{\bf r}\to\infty}\langle{\bf r}|\psi(0)\rangle
    =1-\frac{\mu g}{2\pi\hbar^2}\frac{1}{|{\bf r}|}.
\end{equation}
By definition, the scattering length is the radius of the first node of 
wavefunction at
zero energy. Setting Eq. (\ref{rtoinftypsi0}) equal to zero and solving for 
${\bf r}$ then
gives
\begin{equation}
\label{ga}
    a=\frac{\mu g}{2\pi\hbar^2},
\end{equation}
which relates the scattering length $a$ and the normalization of the 
reference T-matrix,
giving the standard expression $g=2\pi\hbar^2a/\mu$.

\subsection{The s-wave scattering approximation}
\label{lowkApp}

In this section we will discuss the long-wavelength properties of the 
reference T-matrix
(\ref{T'}) and motivate the s-wave scattering approximation in which we 
replace this
T-matrix by its long-wavelength (low energy) limit. From Eq. (\ref{T'}) we 
have
\begin{equation}
    \langle {\bf r}|\hat{T}_{\hat{H}',\hat{V}}|{\bf r'}\rangle=gD({\bf 
r},{\bf r}'),
\label{T'App}
\end{equation}
where the kernel $D({\bf r},{\bf r}')$ is defined as normalized to unity. 
We note that the
exact expression for $D({\bf r},{\bf r}')$ via Eqs. (\ref{TEofVG}) and 
(\ref{G'}) is given
by
\begin{eqnarray}
    g\, D({\bf r},{\bf r}')&=&V({\bf r})\delta({\bf r}-{\bf r}')
    +\left[\frac{-\mu}{2\pi\hbar^2}\right]\frac{V({\bf r})V({\bf 
r}')}{|{\bf r}-{\bf r}'|}
    +\left[\frac{-\mu}{2\pi\hbar^2}\right]^2V({\bf r})V({\bf r}')
    \int d{\bf r}''\frac{V({\bf r}'')}{|{\bf r}-{\bf r}''||{\bf r}''-{\bf 
r}'|}\nonumber\\
    &+&\left[\frac{-\mu}{2\pi\hbar^2}\right]^3V({\bf r})V({\bf r}')
    \int d{\bf r}''\, d{\bf r}'''\frac{V({\bf r}'')V({\bf r}''')}
    {|{\bf r}-{\bf r}''||{\bf r}''-{\bf r}'''||{\bf r}'''-{\bf r}'|}+\ldots,
\label{Drr'expand}
\end{eqnarray}
from which we see that $D({\bf r}',{\bf r})=D({\bf r},{\bf r}')$. We will 
make use of these
expressions in what follows.

At present we are interested in the long-wavelength (low energy) properties 
of this
reference T-matrix, hence we begin by expanding (\ref{T'App}) onto momentum 
eigenstates,
yielding
\begin{equation}
    \langle {\bf k}|\hat{T}_{\hat{H}',\hat{V}}|{\bf k}'\rangle
    =\frac{g}{(2\pi)^3}\int d{\bf r}\, d{\bf r}'\, e^{-i{\bf k}\cdot{\bf 
r}}e^{i{\bf k}'\cdot{\bf r}'}
    D({\bf r},{\bf r}').
\label{TH'Vkk'}
\end{equation}
By applying the gradient operators $\nabla_{\bf k}$ and $\nabla_{\bf k}'$ 
we see that
\begin{eqnarray}
    \nabla_{\bf k}\langle{\bf k}|\hat{T}_{\hat{H}',\hat{V}}|{\bf 
k}'\rangle\big|_{{\bf k},{\bf k}'=0}
    &=&
    -i\frac{g}{(2\pi)^3}\int d{\bf r}\,d{\bf r}'\  {\bf r}\, D({\bf r},{\bf 
r}')\nonumber\\
    &=&-\nabla_{{\bf k}'}\langle{\bf k}|\hat{T}_{\hat{H}',\hat{V}}|{\bf 
k}'\rangle\big|_{{\bf k},{\bf k}'=0}.
\label{nablakT}
\end{eqnarray}
>From Eq. (\ref{Drr'expand}) we see that $\int d{\bf r}' D({\bf r},{\bf 
r}')$ is an even
function of ${\bf r}$ provided only that $V(-{\bf r})=V({\bf r})$. Hence 
the r.h.s of Eq.
(\ref{nablakT}) is the integral of an odd function which leads to
\begin{equation}
    \nabla_{\bf k}\langle{\bf k}|\hat{T}_{\hat{H}',\hat{V}}|{\bf 
k}'\rangle\big|_{{\bf k},{\bf k}'=0}
    =\nabla_{{\bf k}'}\langle{\bf k}|\hat{T}_{\hat{H}',\hat{V}}|{\bf 
k}'\rangle\big|_{{\bf k},{\bf k}'=0}=0.
\label{nablazero}
\end{equation}
The fact that the gradient vanishes at ${\bf k},{\bf k}'=0$ implies that 
this point is
either an extremum or inflection point, i.e. the reference T-matrix is 
topologically flat in
the long wavelength limit. Extending this approach will lead to the result 
that the
curvature in k-space is proportional to the second moment of $D({\bf 
r},{\bf r}')$ so that
we can approximate the T-matrix by its long-wavelength limit provided that 
$|{\bf
k}|R\approx |{\bf k}'|R \ll 1$, where $R$ is roughly the radius of the 
kernel $D({\bf
r},{\bf r}')$. Replacing the reference T-matrix by its long-wavelength 
limit and recalling
that the kernel is normalized to unity gives
\begin{equation}
    \langle{\bf k}|\hat{T}_{\hat{H}',\hat{V}}|{\bf k}'\rangle\approx 
\frac{g}{(2\pi)^3},
\label{Tlowk}
\end{equation}
which then leads to the delta-function approximation
\begin{eqnarray}
    \langle{\bf r}|\hat{T}_{\hat{H}',\hat{V}}|{\bf r}'\rangle
    &\approx& \frac{g}{(2\pi)^6}\int d{\bf k} d{\bf k}'\, e^{i{\bf 
k}\cdot{\bf r}}
    e^{-i{\bf k}'\cdot{\bf r}'}\nonumber\\
    &=&g\delta({\bf r})\delta({\bf r}')
\label{deltaapprox}
\end{eqnarray}
which is the main result of this section.


\end{document}